# Arbitrary geometry electromagnetic spatiotemporal vortices from phase velocity shearing

Jordan M. Adams[1*], Daniel Heligman[1], Rajind Mendis[1] and Josh Wetherington[1]

[1]Riverside Research Institute, 2640 Hibiscus Way, Beavercreek, OH 45431

*Corresponding author: jadams@riversidersearch.org

In fluids, vortices form at boundaries between flows of different velocities. Here, we show that pulsed electromagnetic waves form spatiotemporal vortices when light straddles media of different phase velocities. When the resulting relative time-delay is on the order of the pulse duration of light, spatiotemporal optical vortices (STOVs) are formed. This method allows generating arbitrary geometry spatiotemporal vortices—something impossible with previously demonstrated generation schemes. We provide experimental results showing THz frequency light can be encoded with arbitrary spatiotemporal vortex geometries using simple planar fused filament prints made from a commercial 3D printer. We provide several demonstrations of unique geometries including squares, triangles, ring arrays, and arbitrary curves that cannot be generated with existing techniques. We also show that multiple ring and line vortices can be formed simultaneously, which undergo complex reconnections with propagation. While demonstrated at THz frequencies, this concept is directly applicable to visible and infrared light and our theory offers a way for optimizing STOVs for the frequency range and pulse duration of a given light source. This technique provides a practical and straight-forward method to generate arbitrary electromagnetic spatiotemporal vortices and enables production of complex spatiotemporal phenomena like vortex reconnections.

## 1. Introduction

Optical vortices have been extensively studied for decades [1-3] and used for enhancing applications such as communications [4-7], additive manufacturing [8-10], and particle manipulation [11-13]. Recently, there is a growing interest in spatiotemporal optical vortices (STOVs), where the vortex in a wavepacket is oriented perpendicular to the propagation direction [14,15]. STOVs were first found experimentally in nonlinear pulse propagation [16], but later found to be easily generated in a pulse shaping setup [17]. While a pulse shaper could generate a straight-line vortex, more complicated geometries including toroidal vortices require additional modulation techniques [18]. At the same time, fully arbitrary control of spatiotemporal waves is possible with a multi-mode fiber time-reversal technique [19].

While these methods can generate complex spatiotemporal fields, the techniques are currently only bench-top and require large laboratory setups. In contrast, it was recently found that planar grating devices can be used to sculpt spatiotemporal fields. For example, STOVs with arbitrary tilt angles could be generated [20] and azimuthal polarized wavepackets could be transformed into toroidal vortices under the proper design conditions [21]. Additionally, line STOVs could also be generated by designing a gyromagnetic zero-index material [22]. While promising for various real-world applications, these methods rely on symmetry, gratings, or other complex effects which could potentially limit their ability to generate truly arbitrary geometry spatiotemporal vortices. Additionally, they require fabrication facilities which could limit the ultimate scope of users and as a results limit impact to applications.

Besides visible and infrared light, it has been shown that circular shaped antennas coupled with metasurfaces can be used to generate toroidal STOVs at GHz frequencies [23], where the term STOV is still used despite the vortices occurring in the electromagnetic spectrum outside optical frequencies. In between optical and microwave, tilted line STOVs have been demonstrated in the THz regime using two-color field induced air-plasma filamentation [24]. Additionally, it was shown that pairs of STOVs form in sub-cycle THz pulses that experience a shearing in phase [25].

On the other hand, when including the fourth-dimension of time, electromagnetic waves can behave similarly to waves in fluids. Electromagnetic analogs to fluid wave phenomena like vortex reconnections have been recently demonstrated with both nonlinear and linear propagation of combinations of spatiotemporal vortices [26-28]. Recently, it was also shown that the Kelvin-Helmholtz instability, which forms at the interface between fluids flowing at different velocities [29,30], can also occur with nonlinear electromagnetic pulse propagation [31].

In this paper, we show electromagnetic spatiotemporal vortices form at the boundaries between different phase velocities of light propagation, which share some similarity to Kelvin-Helmholtz vortices forming in fluids with discontinues in flow velocity. From this, we show that arbitrary geometry STOVs in THz frequency waves can be easily generated from simple planar devices printed from commercial fused-filament 3D printers. The device is designed to cause a phase-velocity shear that, by choice of material thickness, culminates in a spatially varying time shift. The material geometry is designed to control the phase-shift at each position and thus control the local time delay. STOVs form at the material discontinuities when the relative time delay is on the order of the pulse duration, where ref. 25 presents a special case of formation with unoptimized time-delay for a single cycle pulse. Single vortices with controllable location can be formed by properly designing the time-shift, while pairs form otherwise. As the STOVs form at the spatial transition between the time-shifted and unaffected field, we can easily and straight-forwardly design the prints to craft arbitrary geometry STOVs—geometries unattainable with previous techniques. Such arbitrary control allows us to generate vortex fields that undergo complex sequences of reconnections. Finally, this allows us to generate vortices that resemble written characters.

We believe this paper will make STOVs available for a wider audience and allow for integration in portable devices, which will ultimately spread the impact to more applications. While demonstrated in the THz or millimeter-wave regime, the concept is directly applicable for generating STOVs in visible and infrared light although will require more precise control over material thickness. Additionally, this work now will empower researchers to easily generate arbitrary geometry vortices. As straight-line STOVs are already being tested for communication applications [32], we believe our method for arbitrary geometry STOVs could be harnessed for further improving high-bandwidth communication. Finally, the results of this work could aid investigations into complex physical processes that occur, such as viscous fluid reconnections or magnetic reconnections.

## 2. Results

2.1 Spatiotemporal Optical Vortex Formation from Time-shifts with Numerical and Analytic Free-Space Propagation

Kelvin-Helmholtz vortices form at the boundary between fluids flowing at different velocities [29]. We now investigative a similar scenario for electromagnetic waves—propagating light that straddles both a material of index $n$ and free-space. One half of the light beam has phase velocity $c/n$ and the other half has phase velocity $c$, with being $c$ is the speed of light. To study this, we use finite-difference time-domain numerical propagation with MEEP [33] to see how light evolves with this phase velocity shear. For this analysis, the input is set to an $x$-polarized plane wave with 0.3 THz center frequency and 0.15 THz the bandwidth. The slab has an index 1.6 and thickness of 0.8 mm. The electric-field results are shown in Figure 1 (a)-(d). The light field gets delayed in the material, and a fork in the electric field appears in the region of the material boundary (Figure 1 (b) and (c)). The phase-fork is evidence of vortex formation, but the spiral phase and intensity null associated with vortices are not well developed until further propagation in free-space (Figure 1 (c)).

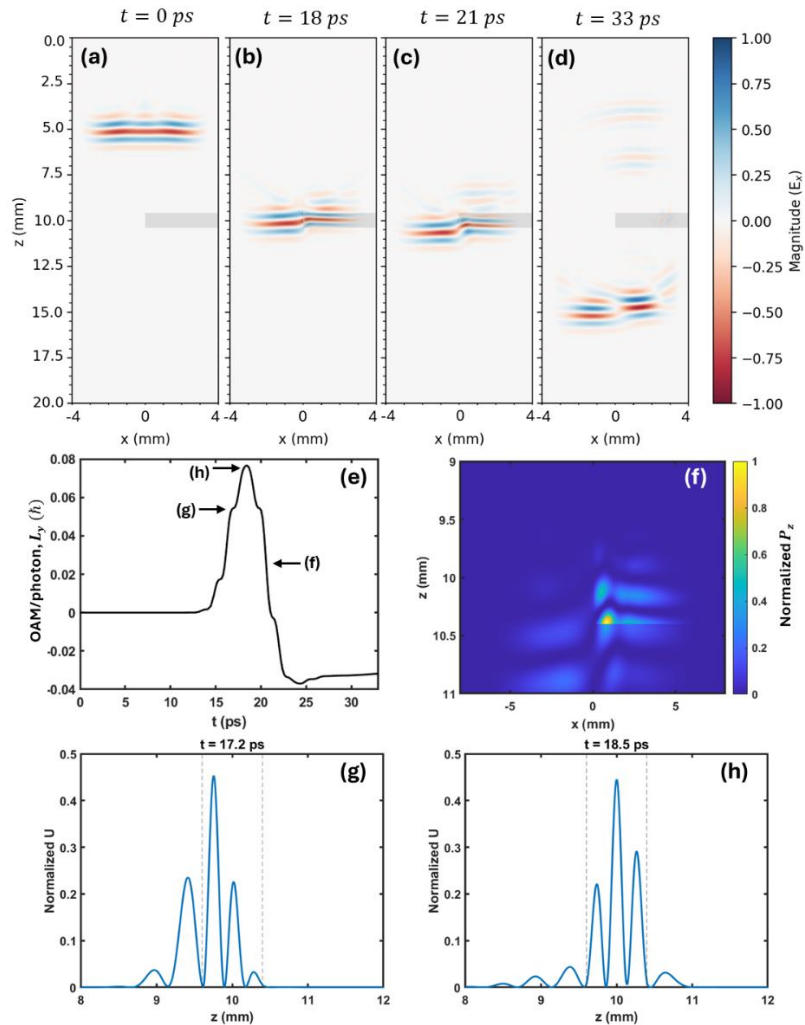

Figure 1: Spatiotemporal vortices form when light straddles and propagates through media with different phase velocities. (a)-(e) The plots show the $x$-polarized electric field found from a finite-difference time-domain simulation of light straddling and propagating through a slab with index n=1.6. The phase velocity discontinuity results in the formation of a spatiotemporal vortex. The signal is attenuated near the boundaries to prevent numerical reflections using an absorbing boundary condition. (e) The time-variation of the total transverse OAM, $L_y$, which demonstrated the field's OAM changes while interacting with the material. (f) The normalized $P_z$ distribution at $t = 20.6$ ps. The normalized energy density at (g) $t = 17.2$ ps and (h) $t = 18.5$ ps at $x = 0.2$ mm, with dashed lines indicating the location of the slab.

Spatiotemporal vortices, like more well-known optical vortices, have orbital angular momentum (OAM). We can calculate the total transverse angular momentum of the field using [34] $L_z(t) = \frac{1}{U_T}\hat{z} \cdot \epsilon \int r \times \vec{E} \times \vec{B} dxdydz$ to understand how the field obtains OAM, where $\epsilon$ is the inhomogeneous distribution of electric permittivity. For comparison, the angular momentum is normalized to the total number of photons, $U_T = \frac{1}{\hbar\omega_0} \int U dxdydz$, where $U = \frac{1}{2}\left(\epsilon_0|E|^2 + \frac{1}{\mu_0}|B|^2\right)$ is the energy density. The results are shown in Figure 1 (e) which shows the OAM increases as the field enters the materials, decreases as it exits, and levels out with net negative OAM after exiting the slab. On top of this, there are fluctuations in the slope, which can be explained by the momentum and energy density fluctuation incident on the surfaces of the slab, oscillating at the carrier frequency. The momentum density, $P_z = \hat{z} \cdot \epsilon \vec{E} \times \vec{B}$, is shown in Figure 1 (f) for $t = 20.6$ ps and the normalized energy density at $x = 0.2$ mm is plotted for $t = 17.2$ ps and $t = 18.5$ ps in Figure 1 (g) and (h). The OAM slope levels out when the fields are zero at the interfaces, while OAM changes rapidly when the field strength at the interfaces is high. When the light enters the media, a spatial asymmetry appears in $P_z$ due to the reduced phase velocity, which contributes to the total OAM via $xP_z$. Based on the specific geometry, this results in net positive angular momentum. However, as the light exits the slab, the asymmetry in momentum density is removed, but the net time delay results in a phase gradient that contributes to total negative transverse angular momentum. In other words, the media results in transient positive transverse angular momentum while inside the material for a positive difference in momentum density, but simultaneously encodes at net negative angular momentum as the field exits for a negative difference in time shift. This shows that light receives transverse OAM when propagating through regions where its phase-velocity has a spatial discontinuity.

The fundamental vortex formation process and any angular momentum transfer to the electromagnetic field must occur while the light straddles the material. While this light-matter interaction is best simulated with numerical FDTD, an analytic model can closely describe the field after the slab and aid understanding and controlling resulting vortex. For this, we assume the effect of the slab is an additional phase, and we start by looking at the result of a material of index, $n$, and thickness, $\Delta z$, that covers half of a Gaussian spatial profile of a wavepacket. Using the Heaviside function, $\theta(x)$, and setting $z = 0$ to the back interface of the material, we can write the field after the wavepacket passes the material as

$$\psi(\omega, x, y, z = 0) = \left(\theta(lx) + exp\left(i2\pi n'\Delta z \frac{\omega}{c}\right)\theta(-lx)\right)\omega \, exp\left(-\frac{x^2 + y^2}{w_o^2} - \frac{(\omega - \omega_0)^2}{w_\omega^2}\right) \tag{1}$$

where $w_o$ is the initial spatial $1/e$ beam radius, $w_\omega$ is half the $1/e$ temporal-frequency bandwidth, $n' = n - 1$ with $n$ being the materials refractive index and $l = \pm 1$ determines whether the material is located in positive or negative $x$ and will correspond to the topological charge sign. Half the beam profile (e.g. $x < 0$ for $l = 1$) gets an additional phase delay in the temporal-frequency domain, which corresponds to a time-delay in the time-domain. Figure 2 (a) shows the temporal field profile of eq. 1 with a half-cycle time delay and Figure 2 (d) shows a spatiotemporal phase cross-section. For this analysis, $\omega$ is in cycles per second. We set , $\omega_0 = 0.3 \, THz$ and $w_\omega = 0.5 \, THz$, which results in single-cycle pulses.

To demonstrate how the field evolves with propagation after encountering the material, we use the angular spectrum method to find the field at a variable $z$ location [3335],

$$\psi(t, x, y, z) = \mathcal{F}^{-1}\left\{\psi(\omega, k_x, k_y, z = 0) exp\left(i2\pi z\sqrt{\left(\frac{\omega}{c}\right)^2 - k_x^2 - k_y^2}\right)\right\}. \tag{2}$$

Of course, the sharp transition that occurs across the beam profile from the material will lead to both propagating and non-propagating fields. Propagating fields occur only when $\left(\frac{\omega}{c}\right)^2 \geq k_x^2 + k_y^2$ and only spatial frequencies below the temporal frequency cut-off will propagate. The sharp spatial phase transition from the material can distribute energy into high-spatial frequencies and, as a result of the cut-off condition, will cause the propagating field to have a sharp spatial-frequency transition in amplitude. Figures 2 (b) and (e) show the propagating field iso-intensity and phase cross-section immediately after passing the material. The cut-off in $k - \omega$ space results in several zero-crossings in space and time, which result in multiple spatiotemporal vortices with propagation. The precursor to the vortices is evident in Figure 1 (e), where a precursor to a vortex pair is highlighted with arrows. As the field propagates (Figure 2 (e) inset, (c), and (f)), the bottom vortex moves outside of the wavepacket, and the top vortex moves toward the center and develops a more circular phase. Additionally, auxiliary vortices appear, highlighted with arrows. Figure 2 (g) shows the iso-intensity profile for farther propagation, which reveals the field has developed a prominent intensity null at the main vortex location, in addition to secondary vortices. However, for smaller initial beam sizes the secondary vortices are no longer present (Figure 2 (h), (k), (i), and (l)). Finally, as the vortex forms locally at the discontinuity, the edge geometry will control the geometry of the vortex.

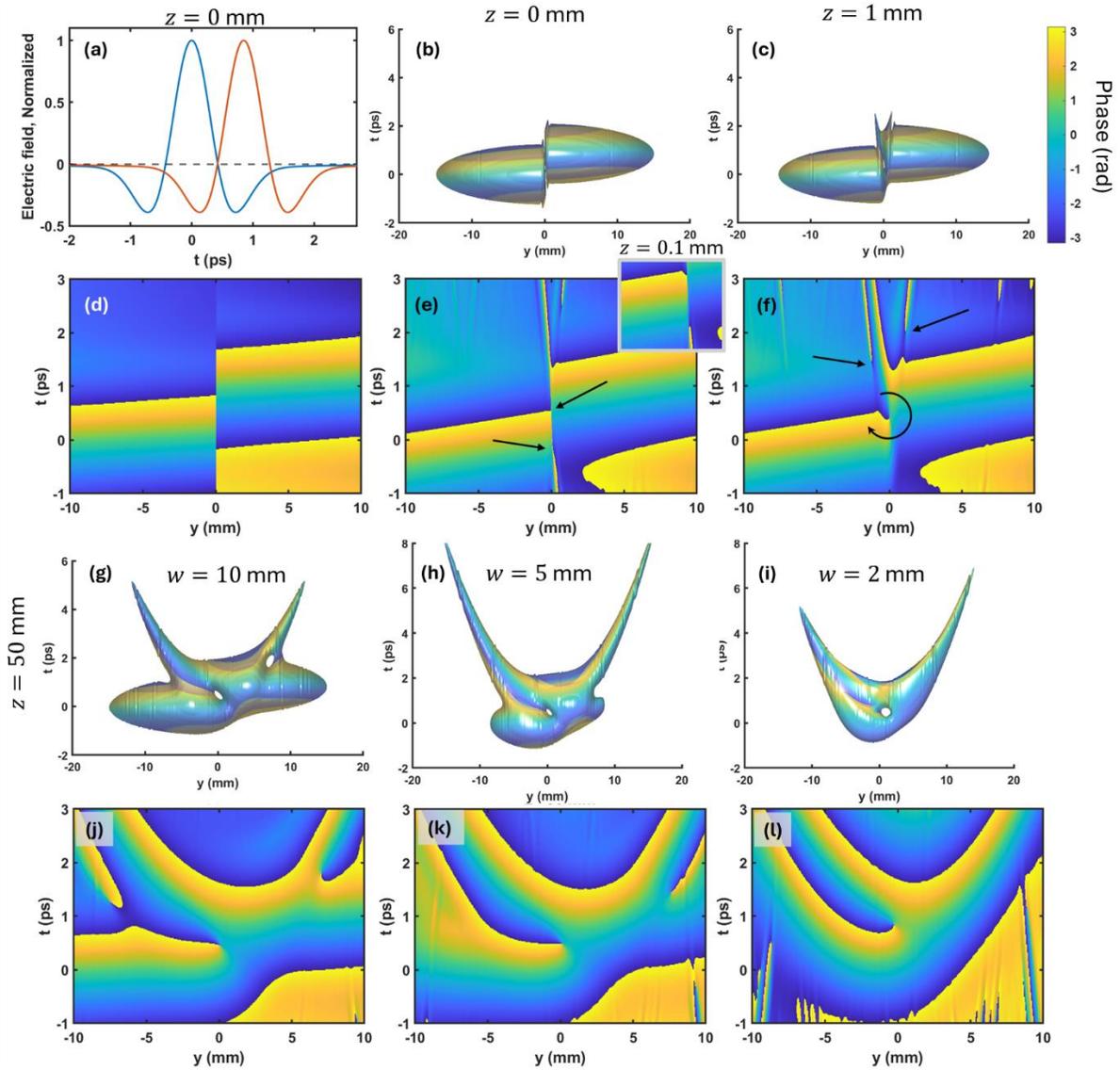

Figure 2: Model simulation results for a straight edge spatially varying time shift, propagated with the angular spectrum method. (a) Iso-intensity plot at 10% intensity with phase-colored surface and (d) phase cross-section of the spatiotemporal field after exiting material, including non-propagating higher spatial-frequencies for an initial beam waist radius of $w = 10$ mm, and temporal-frequency bandwidth parameter $w_\omega = 0.5\ THz$. (b) Iso-intensity and (e) phase cross-section for only the propagating field. Inset on (e) shows how the phase evolves over a small propagation distance. (c) Iso-intensity and (f) phase cross-section at $z = 1$ mm. (g) Iso-intensity and (j) phase cross-section at $z = 50$ mm. (h) Iso-intensity and (k) phase cross-section at $z = 50$ mm for an initial beam waist radius of $w = 5$ mm. (i) Iso-intensity and (l) phase cross-section at $z = 50$ mm for an initial beam waist radius of $w = 2$ mm.

While the vortex is spatially and temporally centered with a half-cycle time shift, deviations in the time-delay result in the vortex moving its location in space and time. Figure 3 (a) and (d) show intensity and phase cross sections for less phase delay, which moves the vortex positive in time and negative in $y$. The vortex moves toward positive in time and in $y$ for larger time shifts (Figure 3 (b), (e), (c) and (f)). In contrast to only one vortex appearing with a half-cycle time shift for single-cycle pulses, longer pulses have multiple vortices appear at the spatial and temporal center for every odd half-cycle multiple or $\frac{2m+1}{2}\frac{\lambda_o}{c}$ time delay, where is $m$ is an integer. This first vortex appearance is shown in Figure 3 (g) and (h) for a time shift near $\frac{1}{2}\frac{\lambda_o}{c}$, while the sixth vortex appears for a $\frac{7}{2}\frac{\lambda_o}{c}$ time shift shown in Figure 3 (i) and (l). Figure 3 (h) and (k) shows no vortex is present at the center for a $\frac{6}{2}\frac{\lambda_o}{c}$ time shift as this an even multiple of $\frac{1}{2}\frac{\lambda_o}{c}$.

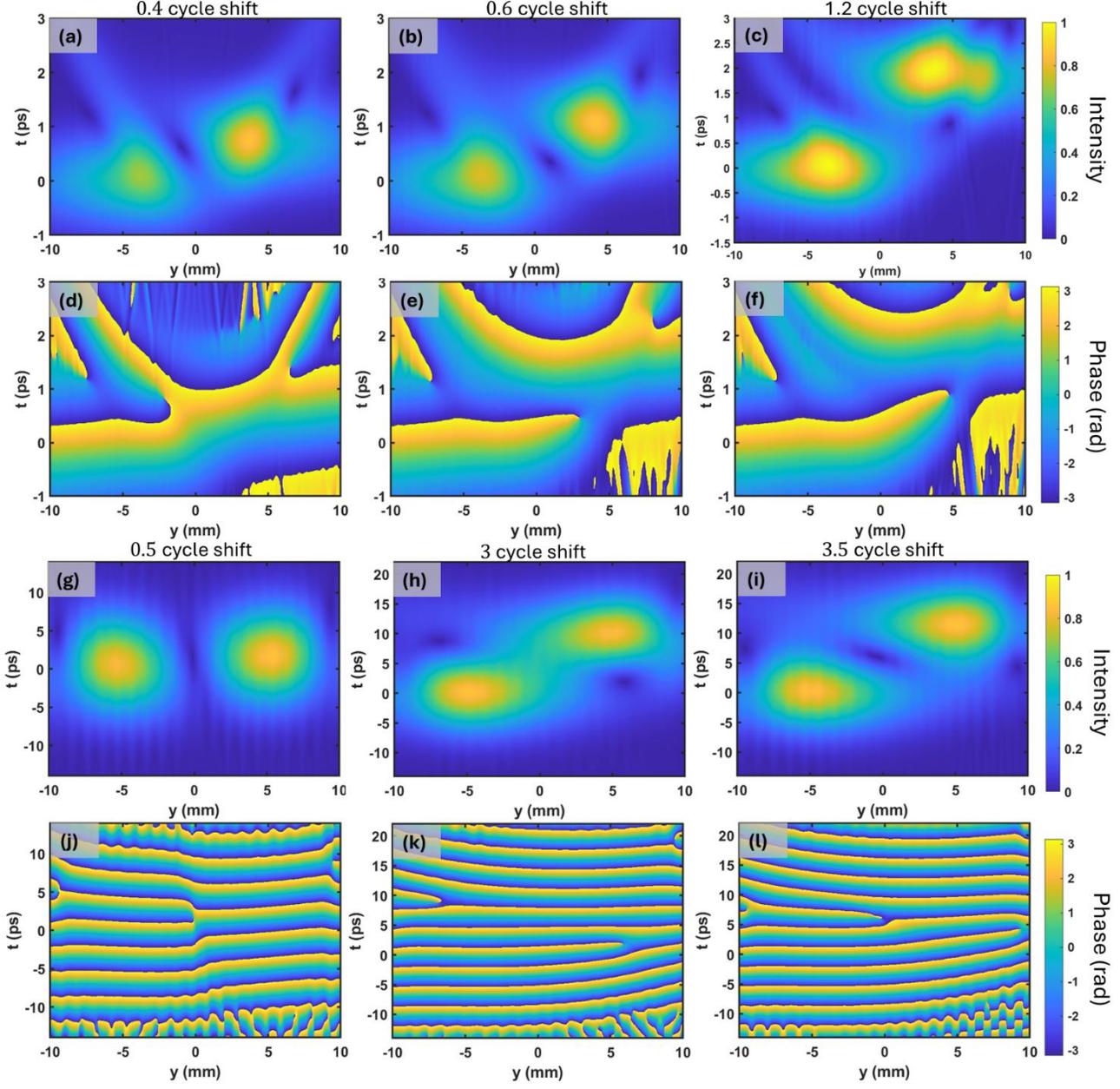

Figure 3: Simulation results for a straight edge spatially varying time shift for variable time-shifts at z = 50 mm. (a) Intensity and (d) phase cross-sections for a 0.4 cycle shift with temporal-frequency bandwidth parameter $w_\omega = 0.5\ THz$. (b) Intensity and (e) phase cross-sections for a 0.6 cycle time-shift. (c) Intensity and (f) phase cross-sections for a 1.2 cycle time-shift. (g) Intensity and (j) phase cross-sections for 0.5 cycle time-time shift, $n'\Delta z/\lambda_0 = 0.5$, with temporal-frequency bandwidth parameter $w_\omega = 0.05\ THz$. (b) Intensity and (e) phase cross-sections for a 3 cycle time-shift. (c) Intensity and (f) phase cross-sections for a 3.5 cycle time-shift.

This behavior can be analytically explained by finding the far-field of eq. 1 while assuming a small bandwidth relative to the center frequency ($k'_x = \frac{x}{\lambda z} \to \frac{x}{\lambda_o z}$)

$$\psi(t,x,y,z) = \left(1 - ierfi\left(\frac{x}{w}\right) + exp\left(-\frac{2n'\Delta z}{c\tau^2}\left(\frac{n'\Delta z}{2c} - t\right) + i2\pi\omega_o\frac{n'\Delta z}{c}\right)\left(1 + ierfi\left(\frac{x}{w}\right)\right)\right)u \qquad (3)$$

where $u = exp(-(x^2 + y^2)/w^2 - t^2/\tau^2)$, $w = \frac{\lambda_o z}{\pi w_o}$ and $\tau = \frac{1}{\pi w_\omega}$. By rearranging eq. 3 and making first order approximations near $t \approx \frac{n'\Delta z}{2c}$ and $x \approx 0$, eq. 3 can be written as

$$\frac{\psi}{Ku} \approx \cos\left(\pi \frac{n'\Delta z}{\lambda_0}\right) + \sin\left(\pi \frac{n'\Delta z}{\lambda_0}\right)\frac{x}{w_x} - i\frac{n'\Delta z}{c\tau^2}\left(t - \frac{n'\Delta z}{2c}\right)\left(\cos\left(\pi \frac{n'\Delta z}{\lambda_0}\right)\frac{x}{w_x} - \sin\left(\pi \frac{n'\Delta z}{\lambda_0}\right)\right) \quad (4)$$

with $K = exp\left(\frac{n'\Delta z}{c\tau^2}\left(t - \frac{n'\Delta z}{2c}\right) + i\pi\omega_o \frac{n'\Delta z}{c}\right)$. For the case of odd multiples of $\lambda_o/2$ phase shifts, $\sin(\pi \frac{n'\Delta z}{\lambda_0})=\pm 1$ and $\cos(\pi \frac{n'\Delta z}{\lambda_0})=0$ which gives

$$\frac{\psi}{Ku} \approx \frac{x}{w_x} + i\frac{n'\Delta z}{c\tau^2}\left(t - \frac{n'\Delta z}{2c}\right) \quad (5)$$

which clearly shows a spatiotemporal vortex centered at $x = 0$ and $t = \frac{n'\Delta z}{2c}$. The circularity of the vortex is maximum when the coefficients on the x and t terms are inversely proportional to the beam width and pulse duration parameters, respectively, $w$ and $\tau$. This means the circularity will be maximum when

$$\frac{n'\Delta z w_x}{c\tau^2} = \frac{w_x}{\tau} \rightarrow \frac{n'\Delta z}{c} = \tau \quad (6)$$

Thus, the circularity is maximum when the time delay reaches the 1/e pulse duration half-width. Of course, a vortex is only present at the beam center when the time delay is odd multiples of $\lambda_0$, $n'\Delta z = (2m + 1)\frac{\lambda_o}{2}, m \in Z$. Equation 3 also shows that shifting the time delay away from this condition shifts the vortices spatially and temporally. One vortex leaves the center while another comes to the center as another odd multiple of $\frac{1}{2}\frac{\lambda_o}{c}$ is met. To have a STOV with optimal circularity centered at the beam center, the time delay must be set to the odd multiple of $\frac{1}{2}\frac{\lambda_o}{c}$ closest to half the 1/e pulse width. As the pulse duration increases and the time shifts remain short, the imaginary term in eq. 5 goes to zero and most of the field is just a first order Hermite-Gaussian beam. Now that we have laid out the theory for STOV formation from spatially varying time shifts, we will present experimental evidence for generating line (section 2.2), toroidal (section 2.3), reconnecting (section 2.4), and arbitrarily shaped vortices (section 2.5).

2.3 Experimental demonstration of line vortices

For the experimental setup, we use a commercial THz time-domain pump-probe system (TeraMetrix™ T-Ray® 5000) which emits single cycle pulses with a baseband frequency range centered around 0.3 THz or a 1 mm vacuum wavelength. With the emitter fixed, the detector is mounted on a raster scanner which measures the time-varying field at each spatial location (Figure 4), allowing direct measurement of the full spatiotemporal electric field at a distance $z$ from the sample. The detector has a 1 mm aperture to sample the spatial profile of the field. To generate the vortices, we used a commercial 3D filament printer (Prusa i3 MK3S) to print a half-circle made of PLA that fully covers half of the emitting 1.5'' diameter lens. The half-circle has four layers which resulted in ~0.8 mm thickness with a refractive index of ~1.6 in the systems frequency range and a time delay of ~$\frac{1}{2}\frac{\lambda_o}{c}$. The sample was taped onto the lens holder and oriented to straddle the beam profile so that the field experiences a straight edge spatially varying time-shift.

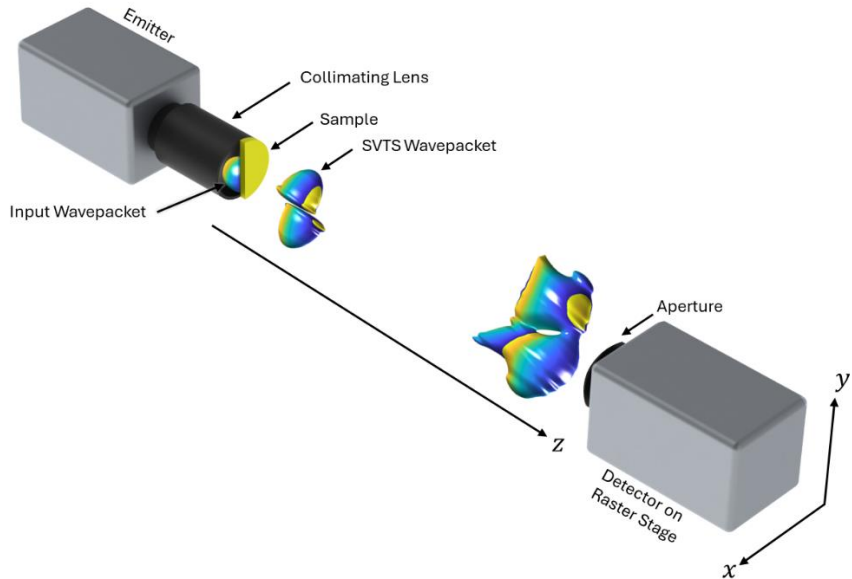

Figure 4: The experimental setup for generating and measuring spatially varying time shift (SVTS) wavepackets. The emitter emits single cycle pulses centered around 0.3 THz. The wavepacket is collimated and fills the 1.5'' diameter emitter lens and the 3D printed sample is attached to the lens holder. The measurement distance is from 3D printed sample to the detector aperture and the detector is mounted on a movable stage for raster-scanning to find the spatiotemporal field. The detector records the instantaneous electric field which enables recovering the spatiotemporally varying intensity and phase.

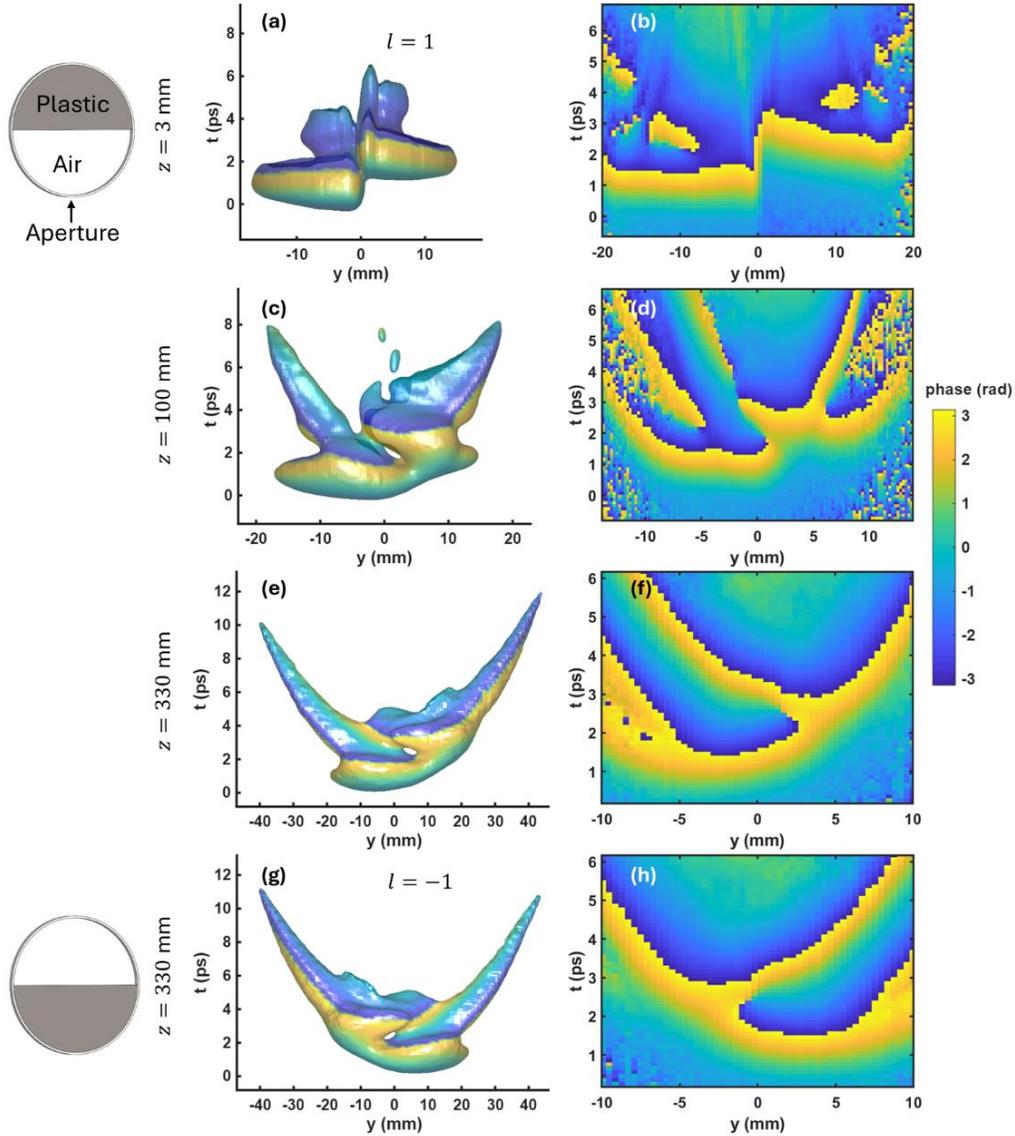

Figure 5: Experimental results for straight edge spatially varying time-shifts. (a) Iso-intensity at 4% intensity with phase-colored surface and (b) phase cross-section for $z=3$ mm for the material covering the top half of the lens ($y > 0$). (c) Iso-intensity and (d) phase cross-section for $z=100$ mm. (e) Iso-intensity and (f) phase cross-section for $z=330$ mm. (g) Iso-intensity and (h) phase cross-section for $z=100$ mm for the material covering the bottom half of the lens ($y < 0$).

The iso-intensity and phase cross-section results are shown in Figure 5 (a), (b), (c), and (d) for a sample to detector propagation distance of $z = 3$ mm , 100 mm, and 330 mm  respectively. The time-shift at $z = 3$ mm transforms into a central spatiotemporal vortex at $z = 100$ mm, with axillary spatiotemporal vortices as predicted in section 2.1. The central spatiotemporal vortex is stable with propagation as demonstrated at $z = 330$ mm. By changing the position of the material to the bottom half of the lens, the opposite topological charge is generated (Figure 5 (g) and (h)).

2.4 Toroidal vortices

While a straight edge straddling the spatial profile of the wavepacket produces line vortices, a circular edge produces toroidal vortices. For this case we printed a disc with a 15 mm diameter which was also 0.8 mm thick. The aperture of the lens holder was covered in Kapton tape (~25 $\mu$m thickness) and the disc was centered and attached to the tape. Figure 6 (a), (d) and (g) show the iso-intensity, raw electric field cross-section, and phase cross-section at distance of 20 mm from the sample. At this point, the field mostly appears to have a circular time-shift, although two vortex ring pairs appear in the phase. At $z = 100$ mm and 330 mm, the vortex is apparent in intensity and the lagging vortex seen at $z =20$ mm is no longer present.

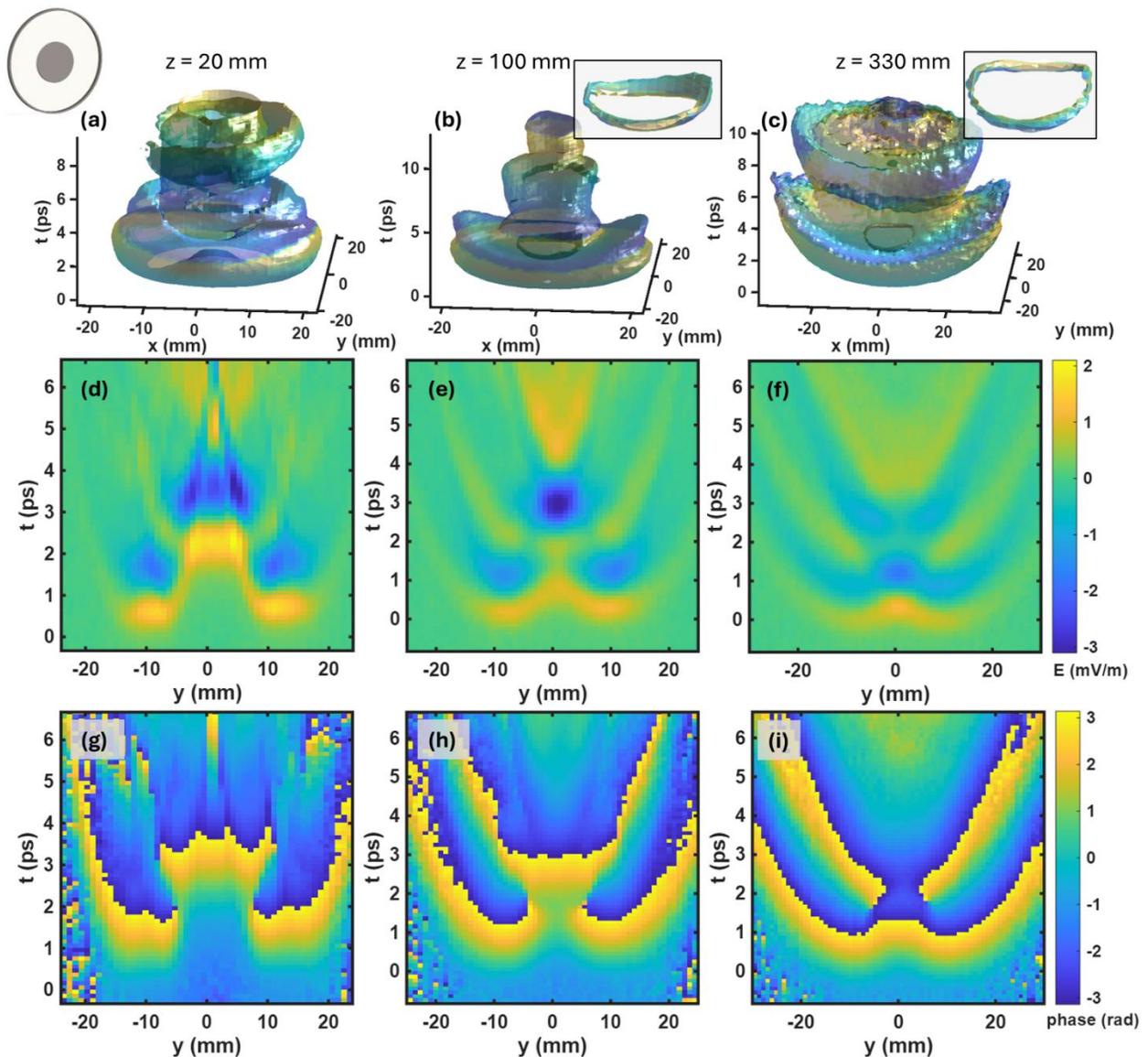

Figure 6: Experimental results for circular spatially varying time-shifts. (a) Iso-intensity at 1% intensity with phase-colored surface, (d) electric-field cross-section, and (g) phase cross-section for $z = 20$ mm. (a) Iso-intensity (d) electric-field cross-section, and (g) phase cross-section for $z = 100$ mm. (a) Iso-intensity, (d) electric-field cross-section, and (g) phase cross-section for $z = 330$ mm.

2.5 Multi-vortex reconnections

If non-parallel vortices are present in the same region, they will reconnect as the field propagates just as fluid vortices reconnect over time [2626-2828]. In this section, we show that SVTS allows for easy creation of multiple vortices and resulting reconnections. First, we show reconnections occurring between two toroidal vortices. To do this, we 3D printed a 0.8mm thick 1.5'' diameter disc with two 5 mm holes with 10 mm center-to-center distance and placed this behind the lens. The iso-intensity plots of the measured field are shown in Figure 7, where we have cropped out the wavepacket envelope to reveal only the internal vortices. Figure 6 (a) shows that two ring vortices are present at $z = 20$ mm. As the field propagates, the rings merge together (Figure 7 (b), $z$=35 mm) and reconnect (Figure 7 (c), $z$=50 mm).

For a 3D print with three equidistant circles, three ring vortices appear (Figure 7 (d), $z = 35$ mm). They simultaneously merge (Figure 7 (e)) and go through a series of reconnections (Figure 7 (f)). While hard to resolve, the process can be explained by Figure 6 (g). The three loops stretch towards each other and reconnect, leaving a closed triangle shaped vortex (Figure 7 (e)). The triangle edges bend toward each other and reconnect, which results in three small rings (Figure 7 (f)). These small rings then merge back together with the original three loops.

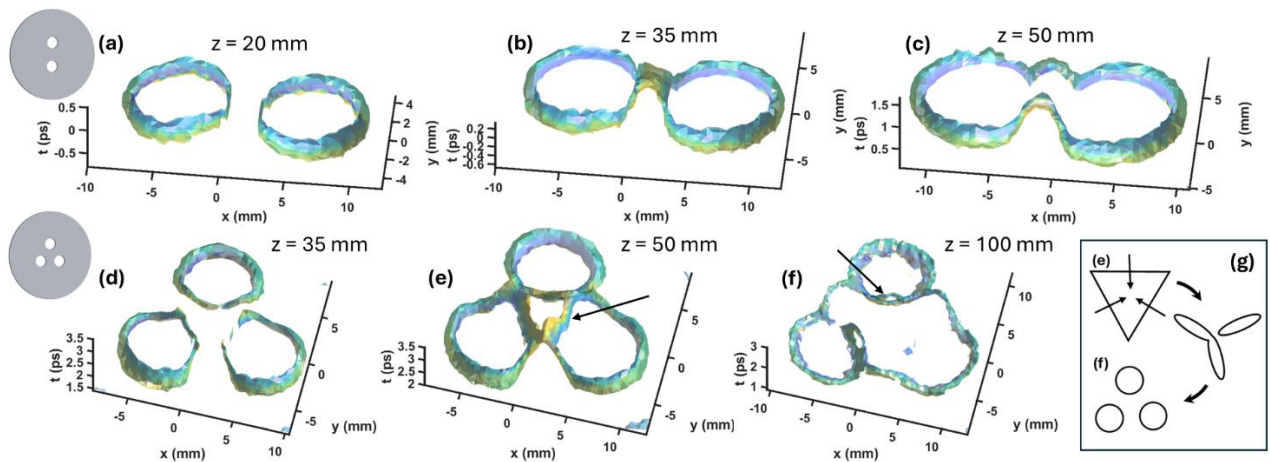

Figure 7: Experimental results for multiple circular spatially varying time shifts resulting in reconnecting loops. (a)-(c) Iso-intensity at 1% iso-intensity with phase-colored surface for $z = 20$ mm, $z = 35$ mm, and $z = 50$ mm for two circular SVTSs. (d)-(f) Iso-intensity at 1% iso-intensity for $z = 20$ mm, $z = 35$ mm, and $z = 50$ mm for three circular SVTSs. (g) Illustration of multi-step reconnection between (e) and (f), with the corresponding small triangle highlighted in (e) with an arrow and one of the small circles highlighted in (f) with an arrow. The extent of the volume was cropped to remove the wavepacket envelope and show only the vortices.

Line vortex reconnections can also be easily generated using SVTS. We 3D printed a base 0.8 mm thick 1.5'' disc, where the first and third quadrants of the disc where given an additional 0.8 mm thickness (Figure 8 (a) inset). Unlike the previous experiments, we use a focusing lens instead of a collimated lens to see the full reconnections occurring through the focus. This would not be possible with a collimating lens since only prefocus effects are observed. The distance $z' = z - f$ is measured relative to the $f = 77$ mm focal plane. Well before the focus, two vortices form with sharp corner geometry (Figure 8 (a) and (d)). Due to the geometry, the second vortex of the vortex pair, discussed in Section 2.1, does not fully disappear on the beam edges and will become important to the field evolution. As the field propagates close to the focus, the two corner shaped vortices split apart and approach each other at two locations forming a loop (Figure 8 (b) and (e)) similar to results in reference [27]. As the field reaches the focus, the secondary vortices of the initial vortex pair also re-merge together at two locations (Figure 8 (c) and (f); Figure 8 (g) and (j)). After the focus, the original corner shaped vortices reconnect, leaving behind a closed loop vortex. This closed loop vortex also reconnects the secondary vortices, resulting now in two new vortices, rotated by ~90° relative to the start.

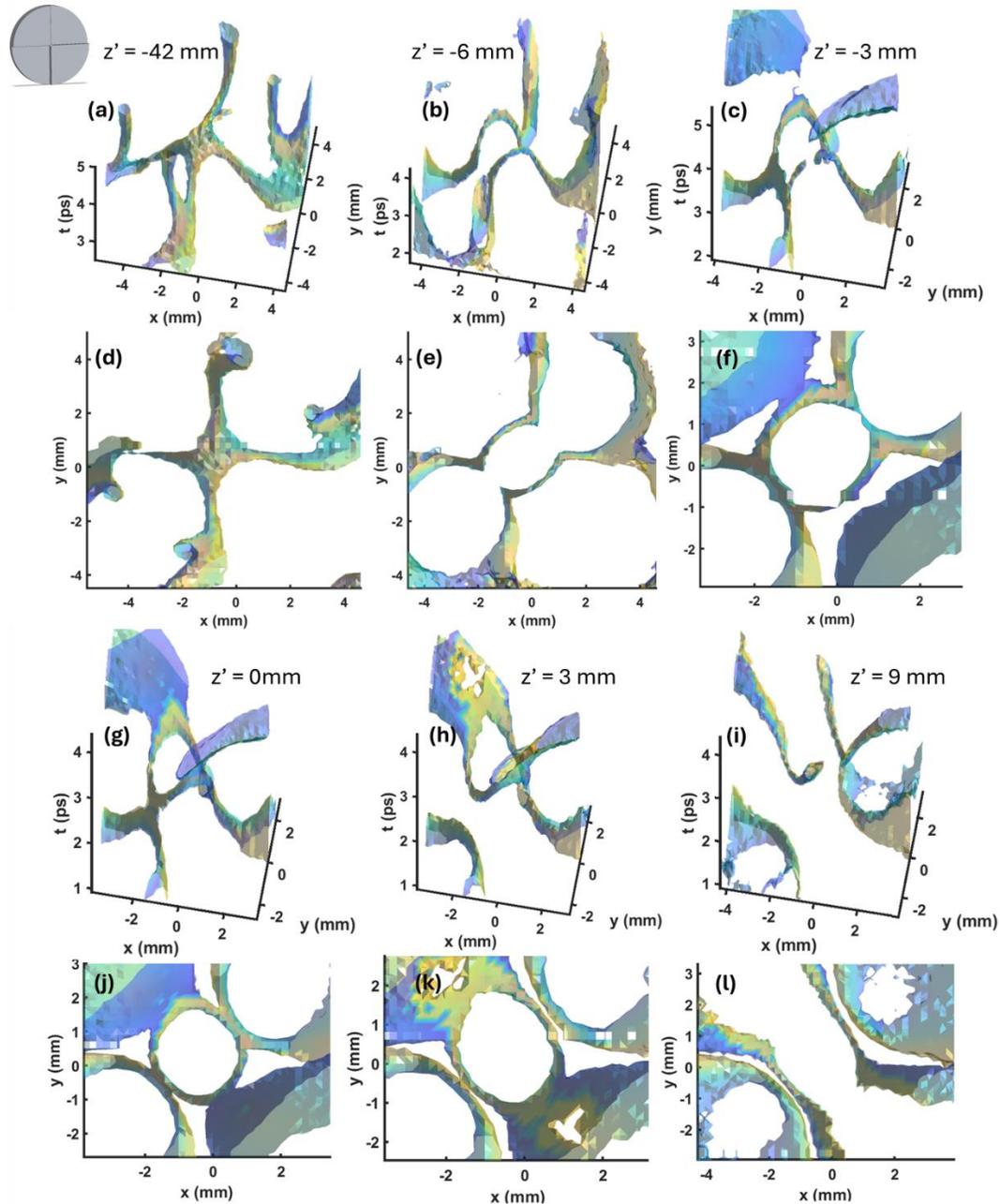

Figure 8: Experimental results for two corner spatially varying time shifts resulting in reconnections occurring through a focus. (a)-(l) Iso-intensity values at 0.7% intensity with phase-colored surface for $z' = $ = -42 mm, -6 mm, -3 mm, 0 mm, 3 mm, and 9 mm from a 3D perspective and a top-view perspective. The extent of the volume was cropped to remove the wavepacket envelope and show only the vortices.

2.6 Arbitrary geometry spatiotemporal vortices

For this last section, we show the generation of arbitrary shaped spatiotemporal vortices. First, we 3D printed a 0.8 mm thick 38 mm disc with a triangle cut-out with 10 mm length sides. The results for propagation with a collimating lens are shown in Figure 9 (a)-(c) for increasing propagation distance. The triangle vortex fully forms and separates from the secondary vortex at $z = 25$ mm. The top view iso-intensity and phase cross-section show a triangle shaped intensity null and $\pi$ phase shift. Second, we 3D printed a square cut-out with 12 mm side-lengths. The results for increasing propagation distance are shown in Figure 9 (d)-(e) . In this case, the vortex forms at $z = 5$ mm and the square intensity null and $\pi$ phase shift are apparent. However, this, along with the triangle shape, quickly becomes more circular with further propagation.

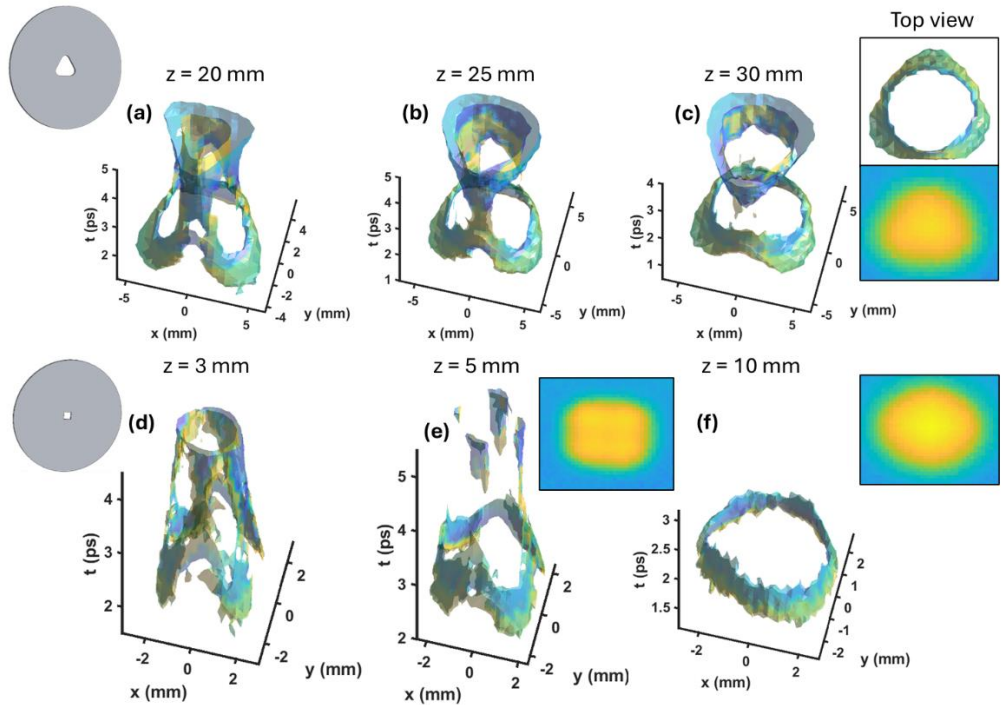

Figure 9: Experimental results for triangle and square spatially varying time shifts. (a)-(c) Iso-intensity plots at 1% intensity with phase-colored surface for $z =$20 mm, 25 mm and 30 mm for a triangle SVTS. Inset on (c) shows a top view of the iso-intensity and time cross-section of the phase. (d)-(f) Iso-intensity plots at 0.4% intensity for $z =$3 mm, 5 mm and 10 mm for a square SVTS. Inset on (e) and (f) shows a time cross-section of the phase. The extent of the volume was cropped to remove the wavepacket envelope and show only the vortices.

Next, we 3D printed a 0.8 mm thick 38 mm disc with the lower-case letters "rri" cut-out, standing for Riverside Research Institute. Figure 10 (a) and (c) show the iso-intensity profile and a phase-cross section at $z = 15$ mm. While the letters clearly form, the resemblance is not perfect as certain features of the "$r$" bend in time instead of spatially. With propagation, the letters reconnect into a single vortex (Figure 10 (b) and (d)).

To obtain better resemblance to the letters, we also generated the letters starting with an array of ring vortices (Figure 11). The array was designed so that the loops reconnect into the desired shape at a certain location. First, an array of small 1.5 mm circular cut-outs were 3D printed to form "rri". Figure 11 (a) –(c) show the iso-intensity results for increasing propagation. An array of ring vortices first form, then begin reconnecting with further propagation. For larger 5 mm holes and larger separation between the holes, the rings reconnect at farther propagation distances (Figure 11 (d)-(f)). In this way it is possible to control the shape at a desired location in space.

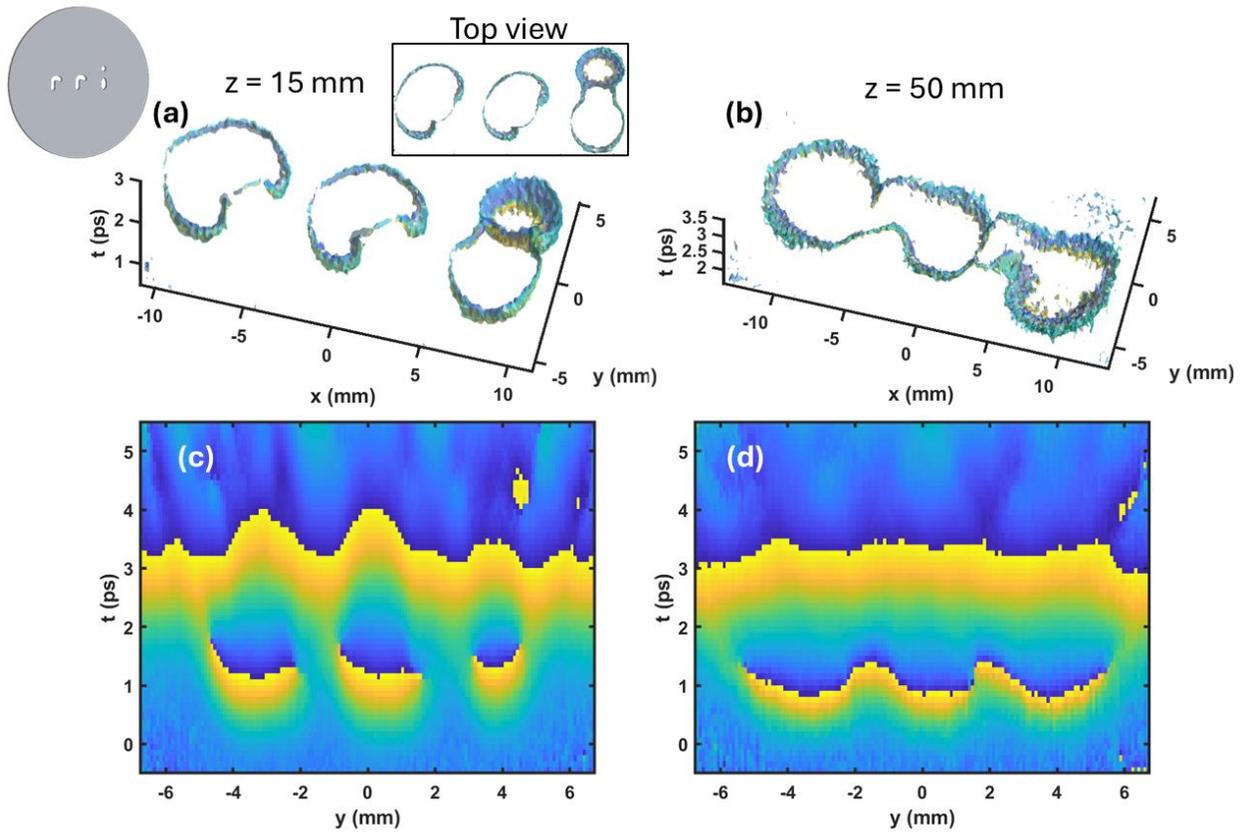

Figure 10: Experimental results for an "rri" spatially varying time-shift. (a) Iso-intensity at 0.2% intensity with phase-colored surface and (c) phase cross-section at $z$ = 15 mm. Inset on (a) shows a top perspective. (b) Iso-intensity and (d) phase cross-section at $z$ = 50 mm. The extent of the volume was cropped to remove the wavepacket envelope and shows only the vortices.

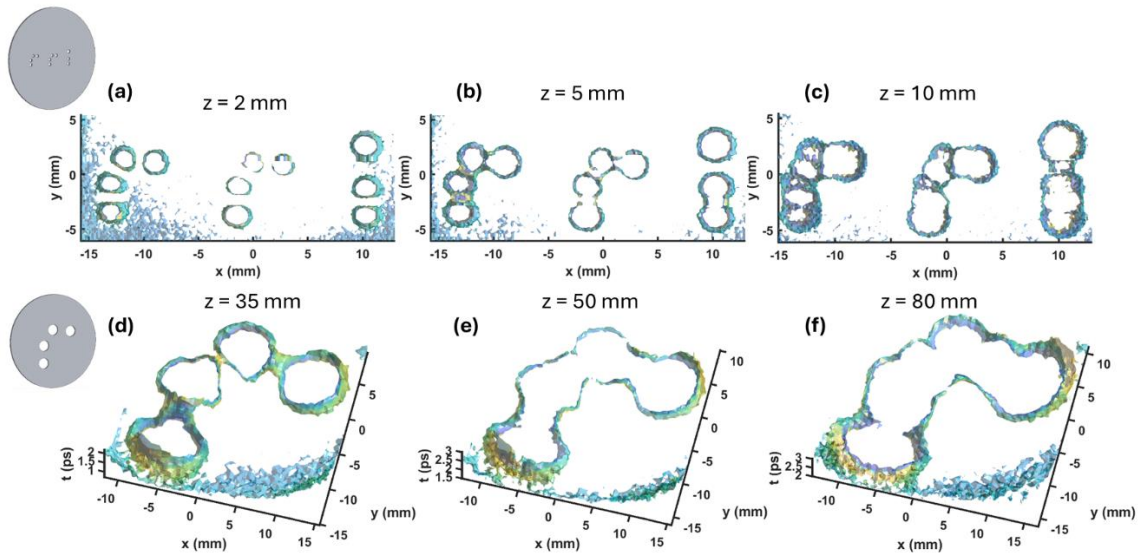

Figure 11: Experimental results for an array of circular spatially varying time-shifts for generating "rri" and "r" spatiotemporal vortices via reconnections. (a)-(c) Iso-intensity plots at 0.4% intensity with phase colored surface at $z$ = 2 mm, 5 mm, and 10 mm for 1.5 mm circular SVTS. (d)-(e) Iso-intensity plots at 0.3 % intensity at $z$ = 35 mm, 50 mm, and 80 mm. The extent of the volume was cropped to remove the wavepacket envelope and show only the vortices.

## 4. Conclusion

In conclusion, we have shown that phase-velocity discontinuities in light produce electromagnetic spatiotemporal vortices. Spatially varying time-shifts can be used to produce arbitrary geometry spatiotemporal vortices. Complex reconnection processes are produced and vortices in the shape of alpha-numeric characters are generated. The light can be preconditioned via arrays of rings to reconnect into desired vortex geometries with further propagation. This ease of design and low-cost fabrication will empower a wider range of researchers to use STOVs for more real-world applications. The light-weight planar device along with the capability for arbitrary STOV design can be adopted for benefiting high-bandwidth free-space communications.

## 5. Methods

The FDTD simulations were performed using the open-source software MEEP. The grid size was 2000x8000, with resolution of 40 points per 100 $\mu$m. The angular spectrum simulation results from eq. 1 and eq. 2 were produced in MATLAB using the fast-Fourier transform (fft) function. The iso-intensity plots for Figure 1 were generated with 51x2001x2001 data points in x,y and t with 100 mm range in x and y and 30 ps range in time. The phase cross-section plots in Figure 1 were generated with 51x6001x4001 data points.

The 3D printed models were designed in SolidWorks, exported as STL files, and sliced using the free online program, Kiri:Moto. The layer height was set to 0.2 mm with 100% infill to ensure all sections of the print resulted in similar time-delays. The models were then printed using the Prusa i3 MK3S 3D printer with SUNLU Pure Yellow polylactic acid (PLA) filament.

The TeraMetrix™ T-Ray® 5000 system captures the raw spatiotemporal electric field. To recover the intensity profile without the fast-time oscillations, we calculated the complex field from the raw real-valued electric field. To obtain the complex spatiotemporal field, the real-valued electric field is Fourier transformed along the time dimension and the field values at negative temporal frequency are set to zero. Then the spatio-spectral profile is inverse Fourier-transformed to obtain the complex spatiotemporal field. The raster step size varied across the measurements between 0.2 mm and 1 mm depending on the overall beam size and feature size of interest. The waveforms-per-pixel averaged sum varied between 30 to 120 depending on the signal strength.

**Research funding:** This work was supported with internal funding from Riverside Research Institute.

**Author contribution:** J.A. performed experiments and analysis. D.H and R.M assisted with experiments. D.H. performed FDTD simulations. J.W advised the project.

**Conflict of interest:** Authors state no conflict of interest.

**Data availability statement:** The data that supports the finding of the current study are available from the corresponding author upon reasonable request.